\documentclass[fleqn,twoside]{article}%
\usepackage{amsmath}
\usepackage{amsfonts}
\usepackage{amssymb}
\usepackage{graphicx}
\usepackage{cite}
\def\be{\begin{equation}}

\def\ee{\end{equation}}

\def\bi{\bibitem}

\begin{document}

\title{Cosmological solutions of the Einstein equation with heat flow.}

\author{Abhik Kumar Sanyal$^1$ and Dipankar Ray$^2$}
\maketitle

\noindent

\begin{center}
$^1$ Department of Physics, Calcutta University, 92 A.P.C.Road, Calcutta-700009, India  \\
$^2$ Department of Applied Mathematics, Calcutta University, 92 A.P.C.Road, Calcutta-700009, India \\
\end{center}

\begin{abstract}
\noindent Cosmological solutions of Einstein's equation for fluids with heat flow in a generalized Robertson-Walker metric are obtained, generalizing the results of Bergmann.
\end{abstract}

\section{Introduction:}

We consider Einstein's equation of general theory of relativity for a fluid with heat flow having the following energy-momentum tensor

\be T^{\alpha\beta} = (\rho + p) v^\alpha v^\beta - pg^{\alpha\beta} + q^\alpha v^\beta + q^\beta v^\alpha,\ee
where, $p$ and $\rho$ are the isotropic pressure and matter density of the fluid respectively, $q_\alpha$ is the heat flux in the radial direction, and $v_\alpha$ is the velocity vector. In the co-moving coordinate system, $v^\alpha = \delta_0^\alpha$, $v_\alpha v^\alpha = -1$ and $q_\alpha v^\alpha = 0$, along with the generalized Robertson-Walker line element

\be ds^2 = A^2 dt^2 - B^2 (dr^2 + r^2 d\theta^2 + r^2 \sin^2\theta d\phi^2),\ee
where $A$ and $B$ are functions of $r$ and $t$. Components of Einstein's equation $R_{\alpha\beta} - {1\over 2} g_{\alpha\beta} R = 8\pi G T_{\alpha\beta}$, had been reduced by Bergmann \cite{1} employing a technique formulated by Glass \cite{2}, to the following single equation,

\be A'' + 2{F'\over F} A' - {F''\over F} A = 0. \ee
In the above, prime denotes differentiation with respect to $x = r^2$, and $F = B^{-1}$. Clearly, one physically relevant assumption is required in order to solve the above differential equation containing a pair of variables $A$ and $F = B^{-1}$. However, a physically meaningful assumption on the metric coefficients $A$ and/ or $B$ is obscure. Bergmann \cite{1} therefore obtained a simple solution under the choice $A = 1$. In this paper, we opt for more general solutions. It is important to mention that once the forms of $A$ and $B$ are known, it is quite trivial to compute the radial component of heat flow, which is given by,

\be q = \left({4r\over G B^2}\right) \left(B\over AB\right)',\ee
where, $G$ is the Newtonian gravitational constant.

\section{Generating solutions:}

\textbf{Case:1.} ~~ $A'' = 0$.\\

\noindent
Under this choice, one obtains

\be A' = Q(t);~~~~~ \mathrm{and}~~~~~ A(x, t) = Q(t) x + P(t).\ee
Thus equation (3) reads as,

\be 2 Q F' - Q x F'' - PF'' = 0.\ee
Integrating the above equation and thereafter dividing throughout by $(Qx + P)^4$, one obtains

\be \left({F\over Q x + P}\right)'  + {h(t)\over (Q x + P)^4} = 0.\ee
Further integration yields,

\be F = {h\over 3 Q} + (Q x + P)^3 L,\ee
and thus,

\be B(x, t) = F^{-1} = \left[{h\over 3 Q} + (Q x + P)^3 L\right]^{-1},\ee
where, $h,~ Q,~ P,~, L$ are all functions of time. Equations (5) and (9) may be used to find explicit form of of the radial component of heat flow $q$, in view of the expression (4).\\

\noindent
\textbf{Case:2.}~~ $A'' \ne 0$.\\

\noindent
Under this choice, $F' \ne 0$, as may be seen from equation (3) and thus one can express $A$ as,

\be A = A(F, t);~~~A' = A_F F';~~~A'' = A_{FF} F'^2 + A_F F'',\ee
where, suffix stands for derivative. So, equation (3) in this case reduces to

\be {A_{FF} + 2{A_F\over F}\over A_F - {A\over F}} d F + {dF'\over F'} = 0.\ee
Integrating the above equation one obtains,

\be \int \left[{A_{FF} + 2{A_F\over F}\over A_F - {A\over F}}\right] d F + \ln{F'} = \ln{\alpha(t)},\ee
or,

\be \exp{\int \left[{A_{FF} + 2{A_F\over F}\over A_F - {A\over F}}~d F\right]} = \alpha(t) {dx\over dF}.\ee
Integrating yet again one obtains,

\be \int\left[\exp{\int \left({A_{FF} + 2{A_F\over F}\over A_F - {A\over F}}~d F\right)} \right] dF = \alpha(t) x + \beta(t).\ee
Therefore, if $A$ is given as a function of $F$ and $t$, then the above integral can be evaluated and hence the solutions may be obtained. Nevertheless, for a particular case, simple solutions may be obtained as follows.\\

Let us consider $F'' = m F$, where $m$ is a function of time alone. So equation (3) may be written as,

\be {U''\over U} = 2{F''\over F} = \pm k^2,~~i.e.,~~ U'' = \pm k^2 U\ee
where, $U = A F$, and $k$ is a function of time. Solutions of the above equation (16) may now be easily found as given below,

\be \begin{split} & U = C_1 e^{kx} + D_1 e^{-kx}, ~~~\mathrm{where,}~m ~\mathrm{is~ positive}~m = k^2,\\&
U = C_1 \cos{(kx)} + D_1 \sin{(kx)}, ~~~\mathrm{where,}~m ~\mathrm{is~ negative}~ m = - k^2,\\&
U = q x + r, ~~~\mathrm{where,}~m=0.\end{split}\ee

\noindent
Subcase-I: $m = k^2$:\\

\noindent
When $m > 0$, equation (15) may be solved to obtain

\be F = C_2 e^{kx\over \sqrt 2} + D_2 e^{-{kx\over \sqrt 2}}.\ee
Now since, $AF = U$ and $B = F^{-1}$, so

\be A = {C_1 e^{kx} + D_1 e^{-kx}\over C_2 e^{kx\over \sqrt 2} + D_2 e^{-{kx\over \sqrt 2}}}; ~~~~B = {1\over C_2 e^{kx\over \sqrt 2} + D_2 e^{-{kx\over \sqrt 2}}},\ee
where, $C_1$, $C_2$, $D_1$, $D_2$ and $k$ are all functions of time. Solution (18) may be used to evaluate $q$ from expression (4).\\

\noindent
Subcase-II: $m = -k^2$:\\

\noindent
When $m < 0$, equation (15) may be solved to obtain

\be F = C_3 \cos{\left(kx\over \sqrt 2\right)} + D_3 \sin{\left(kx\over \sqrt 2\right)},\ee
where, $C_3$ and $D_3$ are functions of time. As before, one can find $A$ and $B$ as,

\be A = {C_1 \cos{\left(kx\right)} + D_1 \sin{\left(kx\right)}\over C_3 \cos{\left(kx\over \sqrt 2\right)} + D_3 \sin{\left(kx\over \sqrt 2\right)}};~~~~B = {1\over C_3 \cos{\left(kx\over \sqrt 2\right)} + D_3 \sin{\left(kx\over \sqrt 2\right)}},\ee
and hence $q$ may be evaluated as well, from the expression (4).\\

\noindent
Subcase-III: $m = 0$:\\

\noindent
In this case $k^2 = 0$, and so equation (15) may be solved to obtain,

\be F = k(t) x + C(t),\ee
which when substituted in equation (3), one obtains

\be kx {d\over dx}\left({dA\over dx}\right) + C{d\over dx}\left({dA\over dx}\right) + 2k \left({dA\over dx}\right) = 0.\ee
Integration yields,

\be A = {f(t) x + g(t) \over k(t) x + C(t)};~~~~~ B = F^{-1} = {1\over k(t) x + C(t)}.\ee
Equation (23) may be used to find the expression for $q$ from equation (3).

\section{Conclusion:}

Summarily, the present paper gives the complete set of cosmological solutions of Einstein's equation with heat flow which was reduced by Bergman to equation (3), either explicitly or implicitly. For $A'' = 0$, solutions have been obtained explicitly and are presented in (5) and (9). For $A'' \ne 0$, on the contrary, solutions are given implicitly by (14). However, some explicit solutions can be obtained for $F'' = +{1\over 2} k^2 F$ as presented in equation (18), $F = - {1\over 2} k^2 F$ as in (20) and $F'' = 0$, as revealed in equation (23). \\

It has already been stated that the solution of equation (3) gives the solution of Einstein's equation for the metric (2) and the energy-momentum tensor (1), where $B = F^{-1}$ and $q$ is the heat flow given by equation (4). Having obtained these solutions, it remains to be shown that these are physically acceptable. Certain energy conditions have to be satisfied, particularly that the energy density is positive everywhere. \\

\noindent
\textbf{Acknowlwdgement:} The authors would like to thank Dr. A. Banerjee for bringing to their atention the work of Bergmann.

\end{document}